\documentclass[12pt]{article}

\usepackage{amssymb,epsfig}
\usepackage{amsmath}

\newcommand{\refeq}[1]{(\ref{#1})}   

\newcommand{\beq}{\begin{equation}}
\newcommand{\eeq}{\end{equation}}
\newcommand{\be}{\begin{eqnarray}}
\newcommand{\ee}{\end{eqnarray}}
\newcommand{\dd}{\mathrm{d}}
\newcommand{\e}{\mathrm{e}}
\newcommand{\mi}{\mathrm{i}}
\newcommand{\de}{\partial}
\newcommand{\vv}{\wedge}

\newcommand{\beh}{\mbox{\boldmath $\eta$}}
\newcommand{\thti}{\tilde{\theta}}
\newcommand{\phti}{\tilde{\phi}}

\newcommand{\nn}{\nonumber \\}
\newcommand{\bbR}{\mathbb{R}}

\DeclareMathOperator{\vol}{Vol}

\DeclareMathOperator{\re}{Re}
\DeclareMathOperator{\im}{Im}
\DeclareMathOperator{\diag}{diag}

\begin{document}


\baselineskip 18pt

\begin{titlepage}

\vfill
\begin{flushright}
\today\\
QMUL-PH-01-07\\
hep-th/0106117\\
\end{flushright}

\vfill

\begin{center}
{\bf \Large Wrapped fivebranes and }\\
\vspace*{3mm}
{\bf \Large ${\cal N}=2$ super Yang--Mills theory}

\vskip 10.mm
{Jerome P. Gauntlett$^{1}$, Nakwoo Kim$^{2}$, Dario Martelli$^{3}$, 
and Daniel Waldram$^{4}$}\\
\vskip 1cm

{\it
Department of Physics\\
Queen Mary, University of London\\
Mile End Rd, London E1 4NS, UK
}\\
\vspace{6pt}

\end{center}
\par

\begin{abstract}
\noindent
We construct $D=10$ supergravity solutions corresponding
to type IIB fivebranes wrapping a two-sphere in a Calabi--Yau
two-fold. These are related in the IR to the large $N$ limit of pure
$\mathcal{N}=2$ $SU(N)$ super Yang--Mills theory. We show
that the singularities in the IR correspond to the wrapped branes
being distributed on a ring. We analyse the dynamics of a probe
fivebrane and show that it incorporates the full perturbative 
structure of the gauge theory. {}For a class of solutions the
two-dimensional moduli space is non-singular and we match the result
for the corresponding slice of the Coulomb branch of the gauge theory.

\end{abstract}

\vfill
\vskip 5mm
\hrule width 5.cm
\vskip 5mm
{\small
\noindent $^1$ E-mail: j.p.gauntlett@qmw.ac.uk \\
\noindent $^2$ E-mail: n.kim@qmw.ac.uk \\
\noindent $^3$ E-mail: d.martelli@qmw.ac.uk \\
\noindent $^3$ E-mail: d.j.waldram@qmw.ac.uk 
}
\end{titlepage}


\section{Introduction}
\label{intro}

It is interesting to generalise the AdS/CFT correspondence~\cite{mald}
to systems with less than maximal supersymmetry and hence richer
dynamics. One strategy is to construct gravity duals corresponding to
branes wrapping supersymmetric cycles as in \cite{malnun}--\cite{hern}. This
approach has been used to construct supergravity solutions corresponding
in the IR to pure ${\cal N}=1$ super Yang--Mills theory in
$D=4$~\cite{malnuntwo}. Here we investigate similar constructions
related to pure ${\cal N}=2$ super Yang--Mills theory in $D=4$.

As in~\cite{malnuntwo}, the super Yang--Mills theory arises from the
IIB little string theory which describes a collection of NS fivebranes in
the limit of vanishing string coupling~\cite{seib}. 
Although this theory is still somewhat mysterious, 
it is known that it reduces at low energies to $D=6$
Yang--Mills theory with 16 supercharges. Dimensionally reducing on a
two-cycle $\Sigma$ should then give a four-dimensional field
theory. In order to preserve supersymmetry, the little string theory
on $\bbR^{1,3}\times \Sigma$ must be coupled to external $R$-symmetry
currents, that is, it must be ``twisted''~\cite{bvs}. This requires
appropriately identifying the $U(1)$ spin connection on the cycle
$\Sigma$ with a $U(1)$ subgroup of the $SO(4)$ $R$-symmetry
group. {}From a geometric standpoint, the twisting is the same as that
arising when a fivebrane wraps a two-cycle $\Sigma$ in a Calabi--Yau
manifold. In particular, the 
$SO(4)$ $R$-symmetry corresponds to the normal bundle to the
fivebrane. The Calabi--Yau condition then requires one to identify a
$U(1)$ factor in the normal bundle with the structure group of the
tangent bundle to
$\Sigma$. In order to preserve eight supercharges, the two-cycle
should be in a Calabi--Yau two-fold. By contrast the twisting
considered in~\cite{malnuntwo} is that associated with fivebranes
wrapping two-cycles in a Calabi--Yau three-fold (with non-generic
normal bundle) and hence preserves four supercharges. As discussed
below, if $\Sigma$ is chosen to be a two-sphere, there are no
additional matter multiplets,
so that at energies smaller than the scale set by
the radius of the sphere we have pure ${\cal N}=2$, $D=4$
super Yang--Mills theory.  

To construct suitable gravity duals we follow the strategy set out in 
\cite{malnun}. We first construct solutions in a truncated $D=7$
gauged supergravity theory and then uplift to 
obtain solutions in $D=10$. The solutions are determined by
two parameters, one of which is the expectation value of
the dilaton. {}For a class of solutions we show that the singularities 
that appear in the IR correspond to the wrapped branes being 
distributed or ``smeared'' over a ring, indicating
which part of the Coulomb branch of ${\cal N}=2$ super Yang--Mills theory
the solutions are related to. We show that the supergravity solution
possesses an $SU(2)$ symmetry corresponding to the $SU(2)$ $R$-symmetry
of the gauge theory. The supergravity solution also possesses a $U(1)$
isometry corresponding to the $U(1)$ $R$-symmetry and we argue that
this is broken to the anomaly free $Z_{4N}$ subgroup by string-worldsheet instantons
as in \cite{malnuntwo}.

To obtain further insight into the solutions we analyse the 
dynamics of a probe fivebrane. 
We find that the dynamics is governed by a holomorphic prepotential on
a two-dimensional moduli space that 
incorporates the exact perturbative effects of the gauge theory. 
In addition, for a class of singular solutions, we remarkably
find a regular moduli space.

Other approaches to studying the large $N$ limit of ${\cal N}=2$ super
Yang--Mills theory have appeared in
\cite{fsone}--\cite{granap}. 
Our solution has some similarities with that of
\cite{pw} on which we will comment at the end of the paper. 

The remainder of the paper is structured as follows. In section 2, we
discuss the twisting of the wrapped little string theory and how it
leads to ${\cal N}=2$ super Yang--Mills theory in the IR. Section 3
presents the supergravity solutions and analyses their
properties. Section 4 analyses the dynamics of the probe fivebranes
and section 5 makes a comparison with gauge theory. Section 6 briefly
concludes.

\section{NS fivebranes wrapped on $S^2$ and ${\cal N}=2$ Yang--Mills
  theory}

Let us start by recalling the IIB little string theory arising from a
collection of flat NS fivebranes. This corresponds to the limit where
the string scale $\alpha'$ is kept fixed while the string coupling
goes to zero~\cite{seib}. In the IR the theory flows to 
six-dimensional Yang--Mills theory. {}For large $N$ the theory
can be studied using supergravity~\cite{imsy,abkt} by analysing the near
horizon limit of $N$ NS-branes\footnote{At some scales the S-dual
D5-brane solutions are more appropriate~\cite{imsy}.} which
is given, with $\alpha'=1$, by
\begin{equation}
\begin{aligned}
   \dd s^2 &= \dd\xi_{1,5}^2 
       + N \left( \dd \rho^2 + \dd \Omega_3^2 \right) , \\
   \e^{-2\Phi} &= \e^{-2\Phi_0} \e^{2\rho} ,
\end{aligned}
\label{eq:flatNS5}
\end{equation}
where $\dd\xi_{1,5}^2$ is the Minkowski metric on $\bbR^{1,5}$ and the
integral of the three-form $H/4\pi^2$ on the three-sphere is $N$.
With regard to supersymmetry, the two chiral $SO(9,1)$ spinors of IIB
string theory each decompose under an $SO(5,1)\times SO(4)$ subgroup as
${\bf 16}_+\to ({\bf 4}_+,{\bf 2}_+)+({\bf 4}_-,{\bf 2}_-)$, where
the subscripts refer to chirality. The $D=10$ Majorana condition
implies that each representation is a symplectic-Majorana spinor
in $D=6$. The fivebrane preserves the ${\bf 2}_+$ part of one 
spinor and the ${\bf 2}_-$ part of the other. In other words, the spin
content of the preserved supersymmetries is the same as that of the
decomposition of a single ${\bf 16}_+$. 

Now suppose the fivebrane is wrapped on a two-sphere. In the IR, at
length scales much larger than the size of the sphere, we have a
four-dimensional field theory. The supersymmetries then have a
further decomposition under $SO(3,1)\times SO(2)\times SO(4)$. 
To preserve eight supercharges we split the $R$-symmetry group via 
$SO(4)\to SO(2)_1\times SO(2)_2$ and identify the $SO(2)$ spin connection 
of $S^2$ with one of the $SO(2)$ factors, say $SO(2)_1$. 
The preserved supersymmetries 
are singlets under the diagonal $SO(2)$ and it is straightforward to
see that this twisting leaves us with eight supercharges or
$\mathcal{N}=2$ in $D=4$. Geometrically it is the same twisting that
arises in the local description of a fivebrane wrapping a sphere
within a Calabi--Yau two-fold. There the $R$ symmetry corresponds to
the symmetry group of the normal bundle. This is naturally split into a
$SO(2)_1$ describing normal directions to the brane within the
Calabi--Yau manifold and an $SO(2)_2$ describing the remaining flat
normal directions. The Calabi--Yau condition, that the two-fold has
$SU(2)$ and not $U(2)$ holonomy, requires the identification of the
first $SO(2)_1=U(1)$ sub-bundle with the $SO(2)$ tangent bundle of the
two-sphere, exactly as discussed above. 

The four scalars in the little string theory transform as a ${\bf 4}$ 
of $SO(4)$ and hence as $({\bf 2},{\bf 1})+({\bf 1},{\bf 2})$ of
$SO(2)_1\times SO(2)_2$. After twisting the former do not have 
any zero-modes on the two-sphere, since the two-sphere is rigid within
the Calabi--Yau manifold, while the latter give rise to two massless
$D=4$ scalars. The components of the gauge field on the two-sphere
have no zero-modes so upon dimensional reduction one has simply a $D=4$
gauge field. These zero modes and their fermionic partners thus
comprise the fields of pure $D=4$, ${\cal N}=2$ $U(1)$ super
Yang--Mills theory. Generalising to $N$ fivebranes we get $SU(N)$ gauge
group. Note that if one were to consider fivebranes wrapped on a genus $g$
Riemann surface, there would be zero-modes arising from deformations
of the two-cycle within the Calabi--Yau manifold, as well as from the gauge
field and one would find $g$ additional hypermultiplets in the 
adjoint representation.

\section{Supergravity solution}
\label{eqsmotion}

The supergravity solutions for the wrapped NS fivebranes only involve
NS fields and hence can be constructed in ${\cal N}=1$ supergravity in
$D=10$. The solutions for wrapped D-fivebranes then follow by S-duality. 
{}Following the strategy set out in~\cite{malnun}, we first construct the
solutions in a truncated theory in $D=7$ and then uplift them to $D=10$. 

A $D=7$, $SO(4)$ gauged supergravity is expected to arise from the
consistent truncation of ${\cal N}=1$ supergravity on a three-sphere
with the $SO(4)$ gauge-fields arising from the isometries of the
sphere. This should be the minimal $SU(2)$ gauged 
supergravity~\cite{paulpvn} coupled to an $SU(2)$ gauge multiplet. 
A Kaluza--Klein ansatz for the bosonic fields was
presented in~\cite{clp} which reduces the bosonic equations of motion
of ${\cal N}=1$ supergravity to $D=7$ field equations. 
We would like to find solutions preserving 1/4 of the supersymmetry. 
However, since we do not have the full reduced
supersymmetric theory\footnote{Note Added: After this work was completed
we became aware of ref.~\cite{salsez} where this $D=7$ gauged supergravity was
constructed.}, we cannot check for supersymmetry directly in
$D=7$. Rather we first construct and solve a set of first-order
seven-dimensional BPS equations, then use \cite{clp} to uplift to give
solutions in $D=10$ and finally directly check that the $D=10$
solutions preserve 1/4 of the supersymmetry both
in ${\cal N}=1$ and type II supergravity. Consequently, logically it is
possible to skip the details of the seven-dimensional derivation given
in the next subsection and start simply with the explicit uplifted
$D=10$ solution given at the beginning of section~\ref{uplift}.

\subsection{The gravity solution in $D=7$}

The bosonic field content of the putative $D=7$ $SO(4)$ gauged
supergravity consists of a metric, $SO(4)$ gauge fields, a three-form and
ten scalar fields parametrised by a symmetric four by four matrix
$T_{ij}$. To construct supergravity solutions 
corresponding to the twistings discussed above we split 
$SO(4)\to SO(2)_1\times SO(2)_2$ and set all gauge-fields
to zero except those corresponding to $SO(2)_1$.
The ansatz for the scalar fields is given by
\be
T_{ij} & = &\e^{y/4}\mathrm{diag}(\e^x,\e^x,\e^{-x},\e^{-x})
\ee
and we set the three-form to zero. With these simplifications, the
equations of motion given in eq. (27) in \cite{clp} can be
obtained from the Lagrangian
\be
{\cal L} & = & \sqrt{g}\left(R -\frac{5}{16}\de_\mu y \de^\mu y - 
\de_\mu x \de^\mu x -\frac{1}{4} \e^{-2x-y/2} F_{\mu\nu} F^{\mu\nu}  
+ 4g^2 \e^{y/2}\right)~.    
\ee 
{}For the metric and gauge field we choose the ansatz
\be
   \dd s^2 & = & 
       \e^{2f(r)}(\dd\xi^2 + \dd r^2) + a^2(r) \dd\Omega_2^2 \nonumber\\
   F & = & \frac{1}{g}\omega_2
\label{twisting}
\ee
where $\dd\xi^2$ is the Minkowski metric on $\bbR^{1,3}$,
$\dd\Omega_2^2=\dd\thti^2+\sin^2\thti\,\dd\phti^2$ and
$\omega_2=\sin\thti\,\dd\thti\wedge\dd\phti$ are the metric element and the
volume form of $S^2$, respectively. Note that as in similar studies it 
is straightforward to also obtain solutions for hyperbolic space
$H^2$, or a quotient thereof. These solutions would be related to
duals of ${\cal N}=2$ gauge theories with matter. The ansatz for the
gauge field can be written in terms of the connection as
$A=g^{-1}\cos\thti\,\dd\phti$, which is proportional to the
spin-connection on the tangent bundle to the sphere, and so
incorporates the desired twisting. The gauge field 
equation of motion is automatically satisfied, whereas the scalar
field equations and Einstein equations give 
\be
    \ddot x + \left(3\dot f + 2 \frac{\dot a}{a}\right) \dot x & = & 
        - \frac{1}{2g^2a^4}\e^{2f-2x-y/2} \nn
    \ddot y + \left(3\dot f + 2 \frac{\dot a}{a}\right) \dot y & = & 
        -\frac{1}{5}\e^{2f} \left( 
            \frac{2}{g^2a^4}\e^{-2x-y/2} + 16g^2\e^{y/2} \right)\nn
    \ddot f + \left(3\dot f + 2 \frac{\dot a}{a}\right) \dot f & = & 
         \frac{1}{10}\e^{2f} \left( 
            \frac{1}{g^2a^4}\e^{-2x-y/2} + 8g^2\e^{y/2} \right) \nn
    4\ddot f + 2 \frac{\ddot a}{a} - 2\dot f \frac{\dot a}{a} & = & 
         \frac{1}{10}\e^{2f} \left( 
            \frac{1}{g^2a^4}\e^{-2x-y/2} + 8g^2\e^{y/2} \right) 
         - \frac{5}{16}\dot y^2 - \dot x^2 \nn
    \frac{\ddot a}{a} + 3 \frac{\dot a}{a}\dot f + \frac{\dot a^2}{a^2} & = & 
         \e^{2f} \left( 
            \frac{1}{a^2} - \frac{2}{5g^2a^4}\e^{-2x-y/2} 
            + \frac{4}{5}g^2\e^{y/2} \right) 
\label{rain}
\ee
with dots denoting derivatives with respect to $r$.

When we uplift to $D=10$, using the formulae in \cite{clp}, to
describe an NS fivebrane configuration we want the warp factor in the
string frame multiplying the unwrapped world-volume directions
$\{\xi^i\}$ to be unity. This leads us to set 
\be
\e^{2f+y/2} & = & 1~,
\label{nowarp}
\ee  
which is consistent with (\ref{rain}).
To find solutions to the remaining equations it is convenient 
to introduce new variables
\be
\e^{2A} & = & \e^{3f} a^2 \nn
\e^{2h} & = & \e^{-2f} a^2~.
\ee
We then find that the equations of motion can be derived from
an effective action whose Lagrangian is given by
\be
L & = & \e^{2A}\left[ 4\dot A^2 - 2 \dot h^2 - \dot x^2 -V \right]\nn
V(h,x) & = & -2\e^{-2h} + \frac{1}{2g^2}\e^{-4h-2x} - 4g^2~.  
\ee
providing that in addition we demand that the Hamiltonian, given by
\be
H & = & \frac{1}{4}\e^{-2A}\left( \frac{1}{4}p_A^2 -\frac{1}{2}p_h^2-
   p_x^2\right) 
+ \e^{2A}V~,
\ee
vanishes. One can then solve the system by finding a set of first-order
BPS equations such that Lagrangian can be written as a sum of squared
terms. That is to say, one can reduce the equations of
motion to a first-order Hamiltonian system, with an associated
Hamilton--Jacobi equation \cite{campos}. By choosing an ansatz for 
the principal function $F=\e^{2A}W(h,x)$ we find that $W$ obeys 
\be
V & = & \frac{1}{4}\left(\frac{1}{2}\de_h W^2 +\de_x W^2 -W^2\right)~.
\label{superp:eq}
\ee
An analytic solution to \refeq{superp:eq} is given by
\be
W & = & -(4g\cosh x+\frac{1}{g}\e^{-2h-x})~.
\ee

The remaining Hamiltonian equations give rise to the following
first-order BPS equations, 
\be
\dot A & = & \frac{1}{4}W = -g\cosh x - \frac{1}{4g}\e^{-2h-x}\nonumber\\
\dot h & = & -\frac{1}{4}\de_h W = -\frac{1}{2g} \e^{-2h-x}\nonumber\\
\dot x & = & -\frac{1}{2} \de_x W = 2g\sinh x -\frac{1}{2g}\e^{-2h-x}~.    
\label{BPS}
\ee
We expect that these equations imply the
solution is supersymmetric. This will be implicitly verified
when we explicitly show that the uplifted solutions in $D=10$ preserve
1/4 of the supersymmetry. 

\begin{figure}[!htb]
\vspace{5mm}
\begin{center}
\epsfig{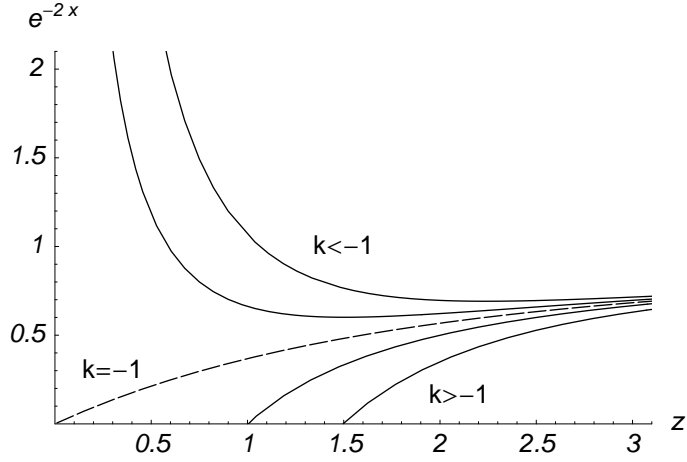}\\
\end{center}
\caption{Behaviour of $\e^{-2x}$ for different values of $k$.}
\label{fig1}
\vspace{5mm}
\end{figure}

We can now solve the BPS equations by using
\be
z = \e^{2h}~,
\ee
as a new radial variable and we find\footnote{For
  negatively curved two-cycles, the solution is $\e^{-2x} = 1 + ( 1 + k
  \e^{2g^2 z} )/2g^2 z$ and $\e^{2A+x}=z\e^{-2g^2z}$.}
\be
\e^{-2x} &=& 1 - \frac{1 + k \e^{-2g^2 z}}{2g^2 z}~, \label{solution}\nn
\e^{2A+x} &=& z \e^{2g^2 z}~,
\label{solution2}
\ee
where $k$ is an integration constant (the other integration constant
can be removed by a coordinate transformation in the
metric~\eqref{twisting}). We will see in the next section that the
parameter $k$ labels different flows from the UV to the IR. In
{}Fig. \ref{fig1} we have plotted the various orbits of $\e^{-2x}$
corresponding to different ranges of $k$. 

\subsection{The supergravity solution in $D=10$}
\label{uplift}

Using the uplifting formulae given in \cite{clp} we can extract the
$D=10$ metric, dilaton and NS three-form. In the string frame our
family of solutions read 
\begin{equation}\label{a}
\begin{aligned}
   \dd s^2 &= \dd \xi^2_{1,3} + z(\dd \tilde{\theta}^2 
           + \sin^2 \tilde{\theta} \dd \tilde{\varphi}^2 ) 
      + g^2\e^{2x}\dd z^2 + \frac{1}{g^2} \dd\theta^2 \\
      &\qquad + \frac{1}{g^2\Omega}\e^{-x}\cos^2 \theta (
           \dd\phi_1 + \cos \tilde\theta \dd\tilde\phi)^2  
      + \frac{1}{g^2\Omega}\e^{x}\sin^2 \theta \dd\phi_2^2~,
\end{aligned}
\end{equation}
where 
\be
\Omega & = & \e^x \cos^2\theta + \e^{-x} \sin^2 \theta~.
\ee
The coordinates $\{\xi^i\}$, $i=0,\dots, 3$, parameterise the 
unwrapped world volume directions, 
$\{\tilde \theta, \tilde\phi\}$ the wrapped ones, 
and $\{\theta,\phi^1,\phi^2\}$ are the angles of the squashed 
and twisted three-sphere. The ranges of the angular coordinates are
explicitly $0\le\thti\le\pi$ and $0\le\phti<2\pi$, while
$0\le\theta\le\pi/2$ and $0\le\phi^1,\phi^2<2\pi$. 
The dilaton is
\be\label{b}
\e^{-2\Phi+2\Phi_0} =\e^{2g^2z}\left(1 -
\sin^2\theta\frac{1+k\e^{-2g^2z}}{2g^2z}  \right)~,
\ee
and the NS three-form reads 
\be\label{c}
   H &=& \frac{2\sin \theta \cos\theta}{g^2\Omega^2} 
        \left(\sin\theta\cos\theta\frac{\dd x}{\dd z}\dd z 
             - \dd \theta \right) \vv 
        ( \dd\phi_1 + \cos\thti \dd \tilde\phi) \vv 
        \dd \phi_2 \nonumber\\
&& + \frac{\e^{-x}\sin^2\theta}{g^2\Omega}\sin
\thti \dd\thti \vv \dd\tilde\phi \vv \dd\phi_2~. \label{Hfield} 
\ee
The solution depends on two parameters: the expectation value of the dilaton
$\Phi_0$ and the parameter $k$ appearing in the function $x(z)$
that is given in (\ref{solution2}) and plotted in {}Fig. \ref{fig1}.

\subsection{Symmetries of the solution}

As solutions of type IIB supergravity, we expect that the 
solution preserves eight supercharges.
To see this we determine the number of Killing spinors
by setting the supersymmetry variations of the fermions 
to zero. {}For vanishing RR fields we require
\be
\delta \lambda &=&  \Gamma^\mu\de_\mu \Phi \tau_3 \beh -
\frac{1}{12}H_{\mu\nu\rho} 
\Gamma^{\mu\nu\rho}\beh = 0 ,\nonumber\\
\delta \psi_\mu &=& \nabla_\mu \beh
-\frac{1}{8}H_{\mu\nu\rho}\Gamma^{\nu\rho}\tau_3\beh = 0~,
\label{gravitino} 
\ee
where $\beh$ is the $SO(2)$-doublet of chiral IIB supersymmetry
parameters and $\tau_3$ is the third Pauli matrix. Using a slightly
non-obvious frame given in the appendix, and inspired by that in
\cite{gkpw}, we find that the Killing spinors are given by
\be
\beh  =  \e^{-\frac{1}{2}(\phi_1\Gamma^{67}+\phi_2\Gamma^{89})}
\beh_0\label{killingspinor}
\ee  
where $\beh_0$ is a constant spinor satisfying 
\be
\Gamma^{6789} \beh_0 &=& -\tau_3 \beh_0 \nonumber\\
\Gamma^{4567} \beh_0 &=& -\beh_0~.
\label{projection}
\ee
This is exactly as desired with eight independent Killing spinors
satisfying the conditions~\eqref{projection}. {}Furthermore, the first
projection is the same as that for a NS fivebrane with tangent
directions $\{0,1,2,3,4,5\}$, while the second is the same as that for
a Killing spinor on a Calabi--Yau two-fold with tangent directions
$\{4,5,6,7\}$. In other words, the supersymmetry preserved 
matches that of a NS fivebrane probe wrapping a two-cycle in
a Calabi--Yau two-fold as expected. Note that as a solution of type I
or type IIA string theory, the solution similarly preserves 1/4 of the
supersymmetry.  

The non-trivial part of the solution 
is six-dimensional and thus is a non-compact example of the class of 
solutions with torsion discussed in \cite{strom}.
In particular the six-dimensional space should be a complex
manifold, with the complex structure $J$ being constructed
from the Killing spinors. {}For our solution, the corresponding two-form
$K$ constructed by lowering one index of $J$ can be written as
\be
K & \equiv & \frac{1}{2}J_{ab} \; e^a \vv e^b
\nonumber\\
&=& e^4 \vv e^5 + e^6 \vv e^7 \pm e^8 \vv e^9 ,
\label{kahlerform}
\ee
where we have directly checked that such a $J$ is indeed integrable.
It will be useful later to note that six-form dual potential $\tilde B$,
defined by 
\be
\dd\tilde B= \e^{-2\Phi}*H,
\ee
is given by
\be\label{calib}
\tilde B= \vol_4\wedge \e^{-2\Phi}K,
\ee
where $\vol_4$ is the volume form on the flat world volume
directions. 

We have seen that the solutions thus preserve the right amount of
supersymmetry to correspond to the supergravity duals of pure ${\cal
  N}=2$ super Yang--Mills theory. Let us now discuss how the $R$-symmetry
of the gauge theory arises as a symmetry of the solution. 
Recall that the classical gauge theory has $SU(2)\times U(1)$ $R$-symmetry
that is broken down to $SU(2)\times Z_{4N}$ by anomalies in the 
quantum theory. The construction we have used, incorporating
the appropriate twistings, automatically has $U(1)\times
U(1)=SO(2)_1\times SO(2)_2$ isometries. These are just shifts in
$\phi_1$ and $\phi_2$. Recall from section two that all 
the fields in the $D=4$, 
$\mathcal{N}=2$ gauge multiplet are singlets under $SO(2)_1$ since
the two-sphere is rigid within the Calabi--Yau manifold. Thus this
first factor should be irrelevant in the IR. However, the scalar
fields transform as a doublet under $SO(2)_2$, 
so we conclude the second factor 
corresponds to the $U(1)$ $R$-symmetry of the classical gauge theory
-- we shall remark later on its breaking down to a $Z_{4N}$ subgroup. 
Actually since our solution includes a round two-sphere, 
corresponding to the cycle on which the fivebrane is
wrapped, the isometry group is larger. If we introduce a  set of
$SO(3)$ left-invariant one-forms in terms of the 
Euler angles $\{\tilde\theta,\tilde\phi,\phi_1\}$
\be
\sigma_1 &=& \quad\cos\phi_1 \dd \tilde\theta + \sin\phi_1\sin\tilde\theta\dd
\tilde\phi\nonumber\\
\sigma_2 &=& -\sin\phi_1 \dd \tilde\theta + \cos\phi_1\sin\tilde\theta\dd
\tilde\phi\nonumber\\
\sigma_3 &=& \dd \phi_1 + \cos\tilde\theta \dd\tilde\phi~,
\label{leftinvariants}
\ee
we see that full isometry group is $SO(3)\times U(1)^2$. (Notice that 
since $0\le\phi_1 <2\pi$, the manifold parametrised by $\sigma_i$ is
topologically $S^3/Z_2$ and hence we have $SO(3)$ symmetry and not $SU(2)$.) 
This ``accidental'' $SO(3)$ isometry would not arise if we were to wrap 
around other two-surfaces. Instead, in general, all that would survive
is a $SO(2)$ symmetry corresponding to rotations of the tangent space
to the cycle. Since this $SO(3)$ cannot provide the remaining $SU(2)$
$R$-symmetry of the ${\cal N}=2$ gauge theory, we see that this
symmetry cannot arise purely from isometries of the solution. Recall
that in the gauge theory the two supercharges form a doublet of the
$SU(2)$ $R$-symmetry. As a consequence we expect to see this symmetry
acting on the Killing spinors of our solution. These can include
transformations which rotate between the two ten-dimensional
chiral IIB spinors. We find that the transformations consistent
with~\eqref{projection} are generated by $\Gamma_{89}$, corresponding
to $U(1)_R$, and also
$\Gamma^{45},\Gamma^{48}\tau_1,\Gamma^{58}\tau_1$, where $\tau_1$ is the first
Pauli matrix, which do indeed
generate $SU(2)_R$.

\subsection{UV and IR behaviour of solutions}

Let us continue our analysis of the solutions by 
examining the asymptotic behaviour of the solutions.
The UV limit is obtained when $z\to \infty$ and hence $x \to 0$.
The metric and
the dilaton are given by
\begin{equation}
\begin{aligned}
   \dd s^2 &=  \dd \xi^2_{1,3}+z\left( \dd\thti^2 + \sin^2\thti
             \dd\phti^2 \right)  
       +  g^2\dd z^2  + \frac{1}{g^2} \dd \theta^2 \\ & \qquad\qquad
       + \frac{1}{g^2}\sin^2\theta \dd\phi_2^2 
       + \frac{1}{g^2}\cos^2\theta \left( 
           \dd\phi_1 + \cos\thti\dd\phti\right)^2 , \\
   \e^{-2\Phi+2\Phi_0} &= \e^{2g^2z} .
\end{aligned} 
\label{asympUV}
\end{equation}
This has the same form as the near horizon limit of the flat
NS-fivebrane solution~\eqref{eq:flatNS5} but with world volume
$\bbR^{1,3}\times S^2$ instead of $\bbR^{1,5}$ and the appropriate twisting. 
Notice that the size of the
$S^2$ is diverging, as in \cite{malnun}, which is related to the fact
that ${\cal N}=2$ Yang--Mills theory is asymptotically free, since on
dimensional reduction from six dimensions the four-dimensional
coupling is inversely proportional to the volume. In addition we 
can connect the seven dimensional gauge coupling $g^2$ to the
number $N$ of wrapped branes via:
\be
\frac{1}{g^2} &=& N~.\label{gN}
\ee
This relation can be seen directly by noting that the integral of
the three-form $H/4\pi^2$ over the transverse three-sphere, which gives the number $N$
of NS-fivebranes, is equal to $g^{-2}$. 
This will be useful later, when comparing with dual gauge theory expectations.

\begin{figure}[!htb]
\vspace{5mm}
\begin{center}
\epsfig{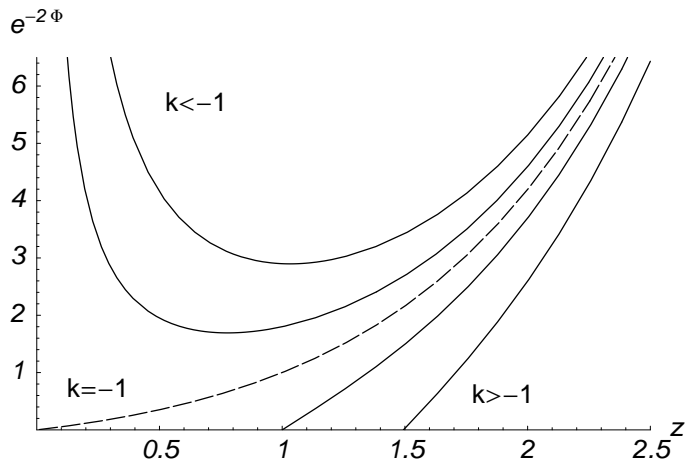}\\
\end{center}
\caption{Behaviour of the dilaton at $\theta=\pi/2$.}
\label{fig2}
\vspace{5mm}
\end{figure}

The one parameter family of solutions, specified by $k$, are all
singular in the IR. {}For $k\geq-1$ the range of the radial variable $z$
is $z_0\le z\le \infty$ where $z_0$ is the solution
of $\e^{-2x(z_0)}=0$ (see Fig.~\ref{fig1}). When $k=-1$ we have
$z_0=0$. {}For these values of $k$ the solutions are singular when
$z=z_0$ and $\theta=\pi/2$: this can be seen from the behaviour of the
dilaton, for example. On the other hand when $k<-1$ we have $0\le z\le
\infty$ and the solutions are singular when $z\to 0$ for generic
$\theta$. We have plotted the behaviour of the dilaton for
$\theta=\pi/2$ in {}Fig.~\ref{fig2}.  

We would like to argue that the $k\ge -1$ 
solutions are related to gravity dual descriptions
of part of the Coulomb branch of ${\cal N}=2$ Yang--Mills theory, 
while the $k< -1$ solutions appear to be unphysical. 
The first piece of evidence for this is that
the singularities are ``good'' for $k\ge -1$ and ``bad'' for $k<
-1$, using the criteria of \cite{malnun}. Recall that the criteria is
based on the behaviour of the norm of the time-like Killing vector
field in the Einstein frame. If the norm is decreasing as one
approaches the singularity, as it is for $k\ge -1$, fixed
proper-energy excitations in the geometry correspond to smaller and
smaller energy excitations in a possible dual field theory, which is
consistent with the singularity corresponding to the far IR
physics. {}For $k< -1$ this norm increases and the singularity would
need excising in some way if one was to develop a similar
interpretation\footnote{Note there are solutions with  ``bad'' singularities
that ultimately get resolved ,e.g., \cite{malnun}. 
However, unlike that example there is no reason to expect this to occur here.}.

A second piece of evidence that $k\geq-1$ corresponds to the Coulomb
branch comes from analysing the metric near the
singularity at $z=z_0$ and $\theta=\pi/2$. We have
$\e^{-2x}\approx 2g^2(z-z_0)$, and defining new variables $y$ and
$\psi$ by  
\begin{equation}
   \begin{aligned}
      \sqrt{2g^2(z-z_0)} &\equiv \sqrt{2gy}\sin(\psi/2) \\
      \theta-\pi/2 &\equiv \sqrt{2gy}\cos(\psi/2)
   \end{aligned}
\end{equation}
with $0\leq\psi<\pi$ and $y\geq 0$ so that $z\geq z_0$ and 
$\theta\leq\pi/2$, we find
\begin{equation}
\begin{aligned}
   \dd s^2_6 &= z_0 \left( \dd\tilde{\theta}^2 + 
           \sin^2\tilde{\theta} \dd\tilde{\phi}^2 \right) \\
      &\qquad + (2gy)^{-1} \left[ \dd y^2 + y^2 \left( \dd\psi^2 
           + \sin^2\psi(\dd\phi_1+\cos\tilde{\theta}\dd\tilde{\phi})^2 \right)
           + g^{-2}\dd\phi_2^2 \right] \\
   \e^{-2\Phi+2\Phi_0} &= 2gy\, \e^{2g^2z_0}
\end{aligned}
\end{equation}
with the singularity located at $y\to 0$. We see that the metric has 
precisely the form of the near-horizon limit of $g^{-2}=N$ NS
5-branes smeared on a circle in the $\phi_2$ direction. 
The fact that the coordinate direction $\phi_2$ is singled out is
supported by our construction. Recall that the twisting we implemented 
corresponds to a NS fivebrane wrapping a two-cycle in a Calabi--Yau
two-fold. Roughly, a radial direction built from $z$ and $\theta$ and 
the angle $\phi_1$ correspond to the two directions transverse to the 
fivebrane and inside the Calabi--Yau two-fold. A different radial
coordinate together with the angle $\phi_2$ correspond to the two directions
transverse to the fivebrane and also to the Calabi--Yau two-fold. (In
fact, this can be made exact by an explicit change of variables
similar to that given in~\cite{gkpw}.) {}From the discussion in section
two, it is the latter directions which are related to the Coulomb branch of
vacua. It appears that the solutions correspond to a slice of the
Coulomb branch where the branes are smeared in a ring parametrised by
$\phi_2$. {}For $k=-1$ it is less clear how to make this direct
argument: the volume of the two sphere that the fivebrane wraps is
shrinking to zero size at the singularity and it is thus difficult to
compare to the smeared fivebrane solution. Our interpretation is that
the $k=-1$ solution corresponds to the smallest radius, namely zero,
on which the fivebranes can be distributed. 
{}Further confirmation of this picture will be developed in the next section
when we study the dynamics of a fivebrane probe.

\subsection{Anomaly in $U(1)_R$}

We have noted that the supergravity solution has a $U(1)$ isometry, corresponding
to shifts in $\phi_2$,
that can be identified with the $U(1)$ $R$-symmetry of classical 
${\cal N}=2$ Yang--Mills theory. In the quantum gauge theory
only a $Z_{4N}$ subgroup is anomaly free, with a $Z_{2N}$ acting 
on the moduli space of vacua (see e.g. \cite{sun}).
As in \cite{malnuntwo} we expect that the anomaly can only
be seen by going beyond the supergravity approximation and by
incorporating string world-sheet instantons. 
In particular consider a string worldsheet
wrapping the two-sphere parametrised by $\tilde\theta,\tilde\phi$
at $\theta=\pi/2$. The flux of the $B$-field over the 
two sphere is a function of $\phi_2$, and at $\theta=\pi/2$, we have,
by repeating the argument in \cite{malnuntwo},
\be
\frac{1}{2\pi} \int_{\phi_2} B = b-2N\phi_2
\ee
for some constant $b$. This flux appears in the worldsheet instanton
calculation and has period $2\pi$. Identifying this with the field
theory $\theta_{FT}$ angle (which has the same period), we find the
anomaly free $Z_{4N}$ subgroup of the $U(1)_R$-symmetry: shifts in $\phi_2$ by $\pi n/N$
with $1\le n\le 4N$ will not change $\theta_{FT}$. Note that this is $Z_{4N}$
since although $\phi_2$ has period $2\pi$, the fermions pick up a minus sign
(as can be seen, for example, from (\ref{killingspinor})), 
and are only periodic under $\phi_2\to\phi_2+4\pi$.

\subsection{D5 solution}

To decouple the four-dimensional gauge theory it is necessary to go
to scales much smaller than the little string theory mass scale. In
this limit the dilaton becomes large and so we should consider the
S-dual solution corresponding to wrapped D-fivebranes \cite{imsy}. 
The D5 brane solution is obtained via the following transformation rules
\be\label{dfivesol}
\Phi_\mathrm{D5} & = & -\Phi_\mathrm{NS5} , \nn
\dd s^2(\mathrm{D5}) & = & \e^{-\Phi_\mathrm{NS5}}  
   \dd s^2(\mathrm{NS5}) , \nn
C_{(2)} & = & -B ,
\ee
where the quantities on the left hand side refer to the S-dual
dilaton, metric, and RR potentials. Similarly the corresponding
six-form potential dual to $C_{(2)}$ is given by $C_{(6)} = -\tilde
B$, which we will use in the next section.

\section{Probing the solution with fivebranes}

A standard technique for analysing the physics of supergravity solutions is
to study the low-energy dynamics of a probe brane. {}For definiteness,
we will probe the wrapped D5-brane solutions given at the end of the
last section. {}For these solutions it is natural to consider a D5-brane
probe wrapping 
the same two-sphere. Physically, one expects that this corresponds to
pulling out one of the $N$ D5 microscopic constituents and studying
its low-energy dynamics. {}From the dual gauge theory point of view this
process corresponds to Higgsing $SU(N)\to SU(N-1)\times U(1)$, and
therefore we expect to find an effective action corresponding to the
$U(1)$ factor, due to the background of the remaining branes. As we
will see, we do indeed find a two-dimensional moduli space with 
dynamics governed by a holomorphic prepotential that incorporates
the perturbative effects of the gauge theory.  
Moreover for $k>-1$ we find a non-singular moduli moduli space despite
the geometry being singular. {}For $k=-1$ we will see that the kinetic energy
of the probe brane is tending to zero in the IR.

The effective action of a probe D5 brane in string frame reads, using the
conventions of \cite{bpp},
\be\label{probeact}
S  & = & - \mu_5 \int \dd^6 y\; \e^{-\Phi_\mathrm{D5}} 
\sqrt{-\det \left[ G_{ab}+ B_{ab} +
2\pi \alpha'F_{ab} \right]} \nonumber\\
&& \qquad + \mu_5\int  [ \exp (2\pi\alpha' F + B)
\vv \oplus_{n} C_{(n)} ]~,
\ee
where $\mu_5=(2\pi)^{-5}{\alpha'}^{-3}$, $F_{ab}$ is a world volume
Abelian gauge field, $B$ is the NS two-form, $C_{(n)}$ are the RR
$n$-forms and it is understood that ten-dimensional fields are pulled
back to the six-dimensional world volume. {}For the D5-brane solution
(\ref{dfivesol}) we have $B=0$, but non-vanishing $C_{(2)}$ and $C_{(6)}$.

We choose the world-volume to have topology $\bbR^{1,3}\times S^2$ and
fix the reparametrisation invariance by identifying the
world-volume coordinates with $\{\xi^i,\tilde\theta,\tilde\phi\}$ in
spacetime. The four scalar fields $z$, $\theta$, $\phi_1$ and $\phi_2$
are then functions of these coordinates. The dynamics of the fivebrane
that are relevant for our purposes are when the fields are just
dependent on the four-dimensional coordinates $\{\xi^i\}$. To get the full
effective theory we will also consider non-zero
four-dimensional gauge fields. {}First, though, let us suppose
the probe brane is at rest so that $z$, $\theta$, $\phi_1$ and
$\phi_2$ are constants and set $F=0$. With this restriction we find
that, in general there is a non-zero potential arising from the
DBI and the six-form contribution to \refeq{probeact}. The contribution
to the effective action is explicitly
\begin{equation}
   S_{\text{potential}} = 
      - \mu_5 \e^{-2\Phi_0} \int \dd^4\xi\, \dd\thti\, \dd\phti\,
          \Omega\e^{-x}z\e^{2g^2z}\sin\thti \left[ 
             1 - \left( 1+ {\cos^2\theta\over \Omega g^2z\e^x\tan^2\thti}
                \right)^{1/2} \right]~.
\end{equation}
{}For there to be no net force on the probe brane we must be at a
minimum of this potential. We find two loci
\begin{equation}
\begin{aligned}
   \text{I :} & \quad \theta=\pi/2 \qquad \text{ for all $k$}, \\
   \text{II :} & \quad z=z_0 \qquad \text{ for $k\geq-1$},
\end{aligned}
\end{equation}
for which the potential, in fact, vanishes. 
It is straightforward to check that both of these configurations
preserve 1/4 supersymmetry. 

The dimensionality of these moduli spaces
can be determined by considering the kinetic energy terms of the
scalar fields arising from the DBI part of \refeq{probeact}, where we
now let $z$, $\theta$, $\phi_1$ and $\phi_2$ be functions of
$\xi^i$. Writing $\partial_i=\partial/\partial\xi^i$, we have, after
integrating over the two-sphere, on locus I 
\be
 S_{\text{scalar}} = -2\pi\mu_5 \e^{-2\Phi_0} \int\dd^4\xi\, 
     z\e^{2g^2z}\left[
         g^2(\partial{z})^2 + \frac{1}{g^2}(\partial\phi_2)^2 \right]
     \qquad \text{(I)}, 
\label{s1}
\ee
independent of $k$, and on locus II
\be
 S_{\text{scalar}} = -{2\pi\mu_5 \e^{-2\Phi_0}\over g^2}\int\dd^4\xi\, 
     z_0 \e^{2g^2z_0} \left[
          \cos^2\theta (\partial\theta)^2 
          + \sin^2\theta(\partial\phi_2)^2 \right]
     \qquad \text{(II)}.
\label{s2}
\ee
In each case we have a two-dimensional moduli space as we expect for 
the Coulomb branch of Abelian ${\cal N}=2$ super Yang--Mills theory.
The moduli space metric on locus~I using the variable $g^2z=\log w$
becomes 
\be
\dd s^2_{{\cal M}_{\textrm{I}}} &=& 
      \frac{4\pi\mu_5\e^{-2\Phi_0}}{g^4}\log w 
           (\dd w^2+w^2\dd \phi_2^2)~.\label{moduliI}
\ee
while on locus~II using $\zeta=\sin\theta$, and recalling that
$0\leq\theta\leq\pi/2$, we obtain the flat disc 
\be
\dd s^2_{{\cal M}_{\textrm{II}}} &=& R^2(\dd \zeta^2+ \zeta^2\dd
\phi_2^2)~.\label{moduliII}
\ee
of radius $R^2=4\pi\mu_5g^{-2}\e^{-2\Phi_0}z_0\e^{2g^2z_0}$.

When $k>-1$ the two loci intersect along the ring $z=z_0,
\theta=\pi/2$, which have exactly the same radius on each
loci. Thus we see that the union of the two loci have the
topology of a plane, and the tension of the probe is finite on the
ring. In other words, quite remarkably, we obtain a non-singular two
dimensional moduli space despite the singularity located at the ring
in the supergravity solution. 

Recall that the $k=-1$ solution is a limiting solution of the $k>-1$
solutions in the sense that the singularity is still good. {}From the
probe point of view one finds that the locus~II degenerates and the
kinetic energy terms of the probe brane become zero when $z=0$ or equivalently
$w=1$. Note that similar behaviour is also realised by the probe for
the apparently non-physical $k<-1$ solutions. 

To complete the four-dimensional effective probe action, now
consider the gauge fields on the brane. Since we are wrapping a
sphere, the only gauge field zero modes come from $F$ with
components along the $\bbR^{1,3}$ directions of the wrapped
probe. {]For the WZ part of the probe action we need $C_{(2)}$ which
is given by $-B$ of the NS solutions. Starting with (\ref{c}) note
that we can write
\begin{equation}
B=-\dd\left[{\sin^2\theta\over g^2 \e^x\Omega}
(\dd\phi_1+\cos\tilde\theta \dd\tilde\phi)\right]\phi_2
\end{equation}
which respects the periodicity of $\phi_2$ after performing a gauge transformation
on $B$. Keeping terms from both the DBI and the WZ part of the probe
action, after integrating over $S^2$,
we find on locus~I
\begin{equation}
   S_{\text{gauge}} = - \frac{\pi(2\pi\alpha')^2\mu_5}{g^2} 
        \int\dd^4\xi \left( \log w\, F^2 + \phi_2 F\tilde{F} \right)
        \qquad \text{(I)}.
\label{g1}
\end{equation}
Note that for $k=-1$, as we approach $z=0$ the kinetic energy terms of the gauge
fields are dropping to zero. {}For locus~II we find
\begin{equation}
   S_{\text{gauge}} = - \pi(2\pi\alpha')^2\mu_5
        \int\dd^4\xi \left( z_0 F^2 + c_0 F\tilde{F} \right)
        \qquad \text{(II)}, 
\label{g2}
\end{equation}
where $\tilde{F}$ is the Hodge dual of $F$ and $c_0$ is a
constant. Note that these results are independent of the dilaton,
which cancels against the contribution from the determinant.

The $\mathcal{N}=2$ supersymmetry implies that the full action should
have the form
\begin{equation}
   S = \frac{1}{8\pi} \int \dd^4\xi\, \left(
          - \im\tau(u)(\partial u)(\partial \bar u) 
          + \frac{1}{2}\re \left[\tau(u) \left(\mi F^2 + F\tilde{F}\right)\right]
          \right)
\label{U(1)action}
\end{equation}
where $u$ is the complex scalar field in the $\mathcal{N}=2$ vector
multiplet and the Yang--Mills coupling
$\tau(u)\equiv(\Theta_{YM}/2\pi)+\mi(4\pi/g^2_{YM})$ is a holomorphic
function of $u$. Comparing with the expressions for the scalar and
gauge field actions~\eqref{s1}, \eqref{s2}, \eqref{g1} and~\eqref{g2},
using $g^{-2}=N$, and setting $\alpha'=1$ we can identify on
locus~I 
\begin{equation}\label{coolio}
   \tau = \mi \frac{2N}{\pi} \log(u/\Lambda) \qquad \text{(I)},
\end{equation}
where $u=\Lambda w\e^{i\phi_2}$ with
$\Lambda=\sqrt{N}/2\pi\e^{\Phi_0}$.  On locus~II, the complex scalar
is given by $u=\Lambda\e^{g^2z_0}\zeta\e^{i\phi_2}$ we find $\tau$ is
constant, 
\begin{equation}
   \tau = \tau_0 \equiv \mi \frac{2N}{\pi} \log(u_0 e^{i c_0}/\Lambda) \qquad
       \text{(II)} ,
\end{equation}
where $u_0$ is the value of $|u|$ at $z=z_0$. 
Note that the logarithmic behaviour of 
$\tau$ on locus I is very reminiscent of the exact perturbative
behaviour in ${\cal N}=2$ super Yang--Mills theory.
In the next section we will make a more precise comparison.

\section{Comparison with gauge theory}

To complete the comparison of the probe effective action calculated
above with that expected from the dual field theory, let us now derive
the expected form of $\tau(u)$ from gauge theory. This will be a
perturbative calculation and will depend on where exactly we are on
the Coulomb branch moduli space. In particular, we will show that our
result is compatible with being at a slice where the branes are
distributed on a ring at $|u|=u_0$.  

{}Following the discussion in~\cite{bpp}, for $SU(N)$
gauge theory the Coulomb branch moduli space is parametrised by the
$N-1$ independent complex expectation values of the adjoint scalar,
representing the relative positions of the $N$ branes, 
\begin{equation}
   \Phi = \diag (a_1, \dots, a_N) ,
\end{equation}
where for $SU(N)$ we have $\sum_i a_i=0$. {}For generic $\{a_i\}$ the
theory is broken to $U(1)^{N-1}$ and the bosonic low-energy effective
action is given by the generalisation of~\eqref{U(1)action}, namely 
\begin{equation}
   S = \frac{1}{8\pi} \int \dd^4\xi\, \left(
          - \im\tau_{ij}\partial a^i \partial \bar{a}^j
          + \frac{1}{2}\re\left[\tau_{ij} \left(
               \mi F^iF^j + F^i\tilde{F}^j \right)\right]
          \right).
\label{N=2action}
\end{equation}
The couplings $\tau_{ij}$ are given in terms of a holomorphic
prepotential $\mathcal{F}$, 
\begin{equation}
   \tau_{ij} = \frac{\partial^2\mathcal{F}}{\partial a^i\partial a^j} .
\end{equation}
Perturbatively the prepotential is given by 
\begin{equation}
   \mathcal{F} = \frac{\mi}{8\pi} \sum_{i\neq j} (a_i-a_j)^2 
            \log \frac{(a_i-a_j)^2}{\mu^2} ,
\end{equation}
and is one-loop exact. There are in addition, non-perturbative
instanton corrections. However, provided $|a_i-a_j|>O(1/N)$ it was argued
in \cite{bpp} that these corrections vanish in the large $N$ limit.  

Now consider the calculation in the last section 
of the dynamics of the probe brane in our supergravity background. 
In field theory this should correspond to breaking a
$SU(N+1)$ theory to $U(1)^{N-1}\times U(1)$ with the first factor
corresponding to the background and the second to the probe brane. If we
write $u$ for the position of the probe brane we have as in~\cite{bpp} 
\begin{equation}
   \Phi = \diag (u,a_1-u/N,a_2-u/N, \dots, a_N-u/N) .
\end{equation}
In the large $N$ limit we then get
\begin{equation}
   \tau(u) = \frac{\partial^2\mathcal{F}}{\partial u^2}
      = \frac{i}{2\pi}\sum_i \log \frac{(u-a_i)^2}{\mu^2} .
\end{equation}
Since $N$ is large we can  replace the sum by an integral
\begin{equation}
   \tau(u) = \frac{i}{2\pi}\int \dd^2a \rho(a)
        \log \frac{(u-a)^2}{\mu^2} ,
\end{equation}
with the density function $\rho(a)$ normalized by $\int\dd^2
a\rho(a)=N$. 

{}From our analysis of the supergravity solution we expect the field
theory dual should have the branes arrayed on a ring at radius
$|u|=u_0$ so $\rho(a)=(N/2\pi u_0)\delta(|a|-u_0)$. Integrating, this
gives
\begin{equation}
   \tau(u) = \mi \frac{N}{\pi} \log(u/\mu)
        \qquad \text{for $|u|\geq u_0$} ,
\end{equation}
and
\begin{equation}
   \tau(u) = \mi \frac{N}{\pi} \log(u_0/\mu)
        \qquad \text{for $|u|\leq u_0$} .
\end{equation}
We see that, up to an overall normalization factor of two, this matches
precisely the form calculated for a probe brane with $k\geq-1$ in both
locus~I and locus~II. This normalisation presumably corresponds to
a normalisation of the gauge fields.

\section{Discussion}

We have presented exact supergravity duals describing fivebranes
wrapped on a two-sphere that correspond in the IR to a slice of the
Coulomb branch of ${\cal N}=2$ super Yang--Mills theory. We have shown
that the solutions have the appropriate symmetries including the fact
that the $U(1)$ $R$-symmetry is broken to a discrete group by
string world-sheet instantons. We have also shown that the IR
singularities correspond to the wrapped fivebranes being uniformly
distributed on a ring and that the dynamics of a probe fivebrane
incorporates the full perturbative effects expected from the gauge
theory. 

It is interesting to compare our results with those of~\cite{pw},
where a one parameter family of supergravity solution was presented
corresponding to a slice of the Coulomb branch of the ${\cal N}=2^*$
theory. This theory arises from mass deformations of ${\cal N}=4$
super Yang--Mills theory and thus the supergravity solution can be
considered to be made from deformed three-branes. There are a number
of similarities with our solutions. The solutions parametrised by
$\gamma$ in~\cite{pw} should be compared with ours as follows:
$\gamma\leq0$ with $k\geq-1$ and $\gamma>0$ with $k<-1$. The former
solutions appear to be physical while those with $\gamma>0$ do not
(at least as far as being dual to ${\cal N}=2^*$ Yang-Mills theory). The
dynamics of a D3-brane probe studied in \cite{bpp,ejp} found for
$\gamma<0$ that the moduli space similarly had two loci. One of the loci,
labelled locus II, is a flat disk which degenerated to a line segment
for $\gamma=0$. The metric in the supergravity solution is singular
on this locus, corresponding to a distribution of D3-branes over the
disk. This is in contrast to our case where the fivebranes are
distributed on a ring for $k\ge-1$. As in our analysis for $k<-1$, the 
moduli space for the probe-brane is completeley regular for $\gamma<0$ despite
the presence of singularities in the solution.
Another similarity is that the dynamics for $\gamma=0$ has the feature 
that kinetic energy terms of the D3-brane probe are tending to zero
as one approached the singularity. We have observed exactly the same features 
here for $k=-1$. 

It would be interesting if one could find generalised supergravity
solutions corresponding to more general slices of the Coulomb branch
in the gauge theory. In our approach this will require relaxing the
ansatz that we considered. There are several obvious directions, for
example incorporating more scalar fields in the $D=7$ gravity ansatz,
which could then include non-circularly symmetric configurations, but
it is not clear that exact solutions could be found.

By considering the dynamics of a probe fivebrane we have argued that our
supergravity solutions include the full perturbative effects of the gauge theory.
To probe the structure of Seiberg--Witten theory \cite{sw} one would also need to
include instanton effects. As noted in \cite{bpp} in the large
$N$ limit one needs to be considering the physics where the vevs
$|a_i-a_j|$ are smaller than order $1/N$ and it would be interesting
if such effects could be incorporated in a supergravity dual. 


\section*{Acknowledgements}
We thank Juan Maldacena, Joe Polchinski, Alessandro Tomasiello 
and Paul Townsend for discussions. 
JPG thanks the EPSRC for partial support, DW thanks the
Royal Society for support. All authors are supported
in part by PPARC through SPG $\#$613.   

\appendix
\section{Orthonormal Frame}
The supersymmetry preserved by the fivebrane is best demonstrated
in a slightly non-obvious orthonormal frame. We choose
\be
e^i & = & \dd x^i \qquad i=0,\dots,3\nn
e^4 & = & z^{1/2} \dd \tilde\theta\nn
e^5 & = & z^{1/2} \sin\tilde\theta \dd \tilde\phi\nn
e^6 & = & \frac{1}{\Omega^{1/2}} 
(-g\e^{3x/2} \cos\theta \dd z + \frac{\e^{-x/2}}{g} \sin\theta \dd \theta ) \nn
e^7 & = & \frac{\e^{-x/2}}{g\Omega^{\frac{1}{2}}}\cos \theta 
(\dd \phi_1+\cos\tilde\theta\dd\tilde\phi)\nn
e^8 & = & -\frac{\e^{x/2}}{\Omega^{1/2}} 
(g\sin\theta \dd z +\frac{1}{g} \cos\theta \dd \theta ) \nn
e^9 & = & \frac{\e^{x/2}}{g\Omega^{\frac{1}{2}}}\sin \theta \dd \phi_2
\ee
where $\Omega$ and $x(z)$ are given in the main text.
This differs from the obvious orthonormal by including a rotation
between the $z$ and $\theta$ tangent directions. A similar kind
of frame was found to be useful in a related context in \cite{gkpw}.

\end{document}